\documentclass[twocolumn,showpacs,floats,amsmath,amssymb,nofootinbib]{revtex4}
\usepackage{graphicx}
\usepackage{dcolumn}
\usepackage{bm}
\usepackage{amsmath}
\usepackage{float}
\usepackage{color}
\usepackage{amsfonts}
\usepackage{amssymb}
\usepackage[all]{xy}
\begin{document}
\title{Thermodynamically allowed phantom cosmology with viscous fluid}

\author{Miguel Cruz}
\altaffiliation{miguelcruz02@uv.mx}
\affiliation{Facultad de F\'\i sica, Universidad Veracruzana 91000, Xalapa, Veracruz, M\'exico}

\author{Samuel Lepe}
\altaffiliation{samuel.lepe@pucv.cl}
\affiliation{Instituto de F\'{\i}sica, Facultad de Ciencias, Pontificia Universidad Cat\'olica de Valpara\'\i so, Avenida Brasil 4950, Valpara\'\i so, Chile\\}

\author{Sergei D. Odintsov$^{a,b}$}
\altaffiliation{odintsov@ieec.uab.es, odintsov@ice.csic.es}
\affiliation{$^{a}$Institut de Ci\`encies de l'Espai, ICE/CSIC-IEEC, Campus UAB, Carrer de Can Magrans s/n, 08193 Bellaterra (Barcelona), Spain\\
$^{b}$Instituci\'o Catalana de Recerca i Estudis Avan\c{c}ats (ICREA), Barcelona, Spain\\
$^{c}$Tomsk State Pedagogical University, 634061 Tomsk, Russia\\
$^{d}$Int. Lab. for Theoretical Cosmology, Tomsk State University of Control Systems and Radioelectronics (TUSUR), 634050 Tomsk, Russia
}

\date{\today}

\begin{abstract}
In this work we present an analysis of the phantom zone in a causal viscous cosmology from a thermodynamic point of view. In this description we consider a chemical potential and the approach of irreversible processes. We assume a flat universe filled with a single dissipative fluid described by a barotropic equation of state, $p = \omega \rho$. This model allows to construct a negative chemical potential for the phantom regime but also this construction allows us to have positive definite temperature and entropy.
\end{abstract}

\pacs{98.80.-k, 05.70.-a, 04.20.Dw}

\maketitle

\section{Introduction}\label{intro}
The phantom scenario is currently a fact that can be supported by various astronomical observations for distinct models \cite{obs1, obs2, obs3, obs4} (see also the Refs. \cite{de1, de2}, where it was shown that a phantom behavior for the dark energy can be sustained by data), however, although at cosmological level we can think of it as an expanding fluid with parameter state $\omega < -1$, it should be mentioned that its intrinsic nature it is not well understood yet. One question that remains to be answered about this unusual component of the universe is around its behavior at thermodynamic level, among other. In this sense, it is possible to find several works where this topic is addressed, but it turns out that only under certain conditions the phantom regime can be allowed at thermodynamic level, see for instance the Refs. \cite{termo1, termo2, elizalde, termo3, lima1} on this topic, and besides it should be clarified that the subject has always been approached from the perspective of standard cosmology and reversible processes.\\

From this point of view it is concluded that the phantom regime fulfills the condition of a positive entropy provided that a negative chemical potential is conveniently introduced, however there are cases in which the concept of temperature must be reinterpreted since it can be negative \cite{termo3} and this is due to the fact that the phantom is described by a scalar field with negative kinetic term, but for this approach it is well known that some pathologies and instabilities are present \cite{phantom}. Usually, phantom thermodynamics leads to negative entropy but positive temperature or to negative temperature and positive entropy.\\

It is important to point out that at cosmological level the magnitude and sign of the chemical potential has a relevant role on the nature of the dark energy parameter state. Under the description of standard thermodynamics, by considering an adequate entropy condition $S \geqslant 0$, was found that a negative chemical potential allows a phantom behavior ($\omega < -1$), on the other hand, a positive chemical potential restricts the parameter state to take values greater than $-1$ \cite{lima1}. In the context of nuclear physics was found that a transition from a positive chemical potential in the hadronic phase to negative quark-gluon phase, represents a physically viable mechanism to describe properly the quark deconfinement \cite{quark}.\\

From a more general perspective, in Ref. \cite{nojiri} it is shown that in the phantom regime the quantum effects have a very important role, at thermodynamic level the inclusion of these effects leads to a good definition for entropy (but divergent in the singularity like other invariants), but also these effects are able to prevent this final singularity from taking place, so that this phantom phase of the universe seems to be only transient, however, for a positive entropy there is a negative temperature. Despite the results obtained for this phase of the universe must be considered with attention, it must be studied in more detail.\\ 

In Ref. \cite{saridakis} it can be found that by employing a varying equation of state parameter, i.e., $\omega = \omega(a)$, it results that the chemical potential can be arbitrary, however, the temperature can be negative or positive if the cosmic fluid behaves as phantom or quintessence. Once again, the phantom regime faces a negative temperature, which physical meaning is still an open subject (see Ref. \cite{negative}). Nonetheless, according to this result there must be a smooth pass from $T>0$ to $T<0$, which means that at some given time the zero temperature must be reached. According to the third law of thermodynamics the value $T=0$ is impossible to achieve by any physical process, including objects like black holes \cite{wald}.\\
    
Therefore, in order to have a better understanding of the phantom regime, a search for new perspectives beyond standard cosmology has now begun. One proposal that has been widely employed is based on the introduction of bulk viscosity as a mechanism to reduce the kinetic pressure of the cosmological fluid, i.e., along the cosmic evolution dissipative effects are present. Some results show that this kind of viscous matter can drive the accelerated cosmic expansion \cite{viscous}. By considering the Supernovae Ia data in the framework of a causal description for dissipative effects, some recent results revealed that bulk viscosity has a relevant role in the cosmic evolution \cite{viscousdata, viscousdata1}. An interesting proposal can be found in Ref. \cite{mostaghel}, where the cosmic expansion is driven by a scalar field described in the context of dissipative cosmology. This model has the peculiarity of presenting a phantom scenario with no big rip singularity. In Ref. \cite{ccl} it was found that within the causal thermodynamic framework, this viscous matter can also mimic the cosmic expansion driven by ordinary or quintessence matter, however, at thermodynamic level it was found that only the decelerated expansion is consistent with the conditions that must be satisfied by the entropy ($dS/dt > 0$ and $d^{2}S/dt^{2} < 0$). Alternatively, in Ref. \cite{scherrer}, it was shown that a viscous cosmology can lead to a phantom behavior considering a modification to the standard expression of the bulk viscosity. A phantom evolution was also studied in Refs. \cite{cataldo, velten} within the viscous cosmological scenarios.\\
 	 
Finally, in Ref. \cite{lepe1} it was shown that under the causal formalism for bulk viscosity the crossing to the phantom zone is allowed when dissipative effects are considered in a cosmological fluid with conserved particle number and when the thermodynamics of irreversible processes is adopted. At thermodynamic level the model verified naturally the requirements for the entropy. In a complementary way, in Ref. \cite{lepe2} it was studied the phantom regime obtained in a non-linear extension of the model given in \cite{lepe1}. In the thermodynamics description of this model it was found that under certain conditions the model is in disagreement with the second law of thermodynamics and additionally the conditions for the entropy are not always fulfilled. This is a clear indication that a large effort must be maintained in order to have an improved understanding of the thermodynamic properties of the phantom regime.\\                
 
This paper is organized as follows: in Section \ref{sec:IStheory} we present some generalities of the phantom solution obtained for some viscous cosmologies. We focus on the causal thermodynamic formalism and we define some important quantities written in terms of this phantom solution. In Section \ref{sec:thermo} we introduce the irreversible thermodynamics description and we show that under this approach it is possible to obtain a negative chemical potential. In Section \ref{sec:temp} we discuss the positivity of the temperature and the entropy once that the phantom solution is taken into account. Finally, in Section \ref{sec:elfinal} we write the conclusions for our work. $8\pi G = c = 1$ units are considered throughout this work.    

\section{Viscous phantom solution from Israel-Stewart theory}
\label{sec:IStheory}
In this section we will provide a general description of the phantom solution found in Ref. \cite{cataldo} within the context of viscous cosmologies, which can be written as
\begin{equation}
H(t) = A(t_{s}-t)^{-1},
\label{eq:phantom}
\end{equation}
where $t_{s}$ is a finite time in the future at which the big rip will occur and $A$ must be a positive constant in order to describe an expanding universe. This solution was found in the Eckart theory and in the full Israel-Stewart model, which is a more adequate approach since the causality is preserved. For this type of solution the cosmic fluid can not describe stiff matter, something similar occurs for quintessence, however a phantom behavior is allowed. The solution given in (\ref{eq:phantom}) will represent a physically reasonable big rip solution always that the dark component of the universe be phantom energy. Additionally, this solution is a more general case than the solution found in Ref. \cite{barrow} by Barrow for dissipative universes, i.e., presents a big rip singularity.\\   

By considering an isotropic and homogeneous universe, the field equations in a flat FLRW spacetime for a dissipative fluid are given as follows
\begin{align}
& 3H^{2} = \rho, \label{eq:fried1}\\
& \dot{H} + H^{2} = - \frac{1}{6}\left(\rho + 3p_{eff} \right), \label{eq:fried2}
\end{align} 
where the dot represents a cosmic time derivative, $H := \dot{a}/a$ is the Hubble parameter ($a$ the scale factor) and $p_{eff} := p + \Pi$ is the effective pressure, being $\Pi$ the bulk viscous pressure\footnote{The phantom solution written in Eq. (\ref{eq:phantom}) is obtained for the set of equations given in (\ref{eq:fried1}) and (\ref{eq:fried2}), when the bulk viscous pressure evolution is described by the full transport equation of the Israel-Stewart model expressed in Eq. (\ref{eq:transport}).}. As pointed out in Ref. \cite{maartens2}, due to the symmetries of the FLRW spacetime, only scalar dissipation is possible (absence of energy flux due to heat flow), therefore the dissipation coming from bulk viscosity makes possible the conversion of kinetic energy of the particles into heat, which implies a reduction in the effective pressure of the cosmic fluid. This effect is fully characterized by the quantity $\Pi$.\\

We will assume a barotropic equation of state for the pressure and density of the fluid, i.e., $p = \omega \rho$, where the parameter state is restricted to the interval $0 \leq \omega < 1$. It is important to point out that the bulk viscous pressure has direct incidence on the $\omega$-parameter, since bulk viscosity is capable to modify the background dynamics \cite{velten}
\begin{equation}
\omega_{eff} = \frac{p + \Pi}{\rho}.
\end{equation}  
By using the Eqs. (\ref{eq:fried1}) and (\ref{eq:fried2}), we can obtain the continuity equation
\begin{equation}
\dot{\rho} + 3H\left(\rho + p_{eff} \right) =  \dot{\rho} + 3H\left[\left(1+\omega \right)\rho + \Pi \right] = 0.
\label{eq:continuity}
\end{equation}
From this last expression and the cosmic time derivative of equation (\ref{eq:fried1}), we can obtain an explicit form for the bulk viscous pressure after a straightforward calculation, yielding
\begin{equation}
\Pi = - \left[2\dot{H}+ 3(1+\omega)H^{2}\right].
\label{eq:pressure}
\end{equation}
In order to have a consistent physical description for the dissipative effects in the expanding fluid, we must consider the full Israel-Stewart transport equation for $\Pi$ \cite{israel}
\begin{equation}
\tau \dot{\Pi} + \Pi = -3\xi H - \frac{1}{2}\tau \Pi \left[3H + \frac{\dot{\tau}}{\tau}-\frac{\dot{\xi}}{\xi}-\frac{\dot{T}}{T} \right],
\label{eq:transport}
\end{equation} 
being $\tau$ the relaxation time, $\xi(\rho)$ is the bulk viscosity and $T$ is the barotropic temperature of the fluid. In the limit $\tau \rightarrow 0$, the non-causal Eckart theory is recovered.\\ 

If we consider as a starting point the equation (\ref{eq:transport}) and assume the standard definition for the bulk viscosity coefficient, $\xi = \xi_{0}\rho^{s}$, being $\xi_{0}$ and $s$ positive constants together with $T = T_{0}\rho^{\omega/(1+\omega)}$, which is obtained by means of the integrability Gibbs condition, where $T_{0}$ is an integration constant and additionally, $\tau = \xi_{0}\rho^{s-1}/[\epsilon (1-\omega^{2})]$ where $0 < \epsilon \leq 1$ for causality conditions and also using the Eqs. (\ref{eq:fried1}), (\ref{eq:continuity}) and (\ref{eq:pressure}), one gets a second order differential equation for the Hubble parameter. 
It is important to point out that the definition given before for the relaxation time, $\tau$, is a more consistent expression since is derived from the speed of bulk perturbations (see Ref. \cite{maartens2}), therefore the differential equation obtained for $H$ with this definition of $\tau$ differs from the one obtained in Refs. \cite{cataldo, israel, maartens}.\\ 

In Ref. \cite{lepe1} it was found that under the election $s = 1/2$, the phantom solution (\ref{eq:phantom}) when inserted in the second order differential equation for the Hubble parameter results in an algebraic equation for $A$ where $A = A(\omega, \xi_{0})$. For this special value of $s$ the model does not admit a de Sitter solution. The positive solution obtained for $A$ made possible the crossing of the phantom divide through the effective parameter $\omega_{eff}$, i.e., the crossing to the phantom zone is due to the contribution of the dissipative effects in the fluid. Also, in Ref. \cite{lepe2} with the same consideration $s=1/2$, the crossing of the phantom divide was obtained for the effective parameter $\omega_{eff}$ using the solution (\ref{eq:phantom}), but in a non-linear extension of the Israel-Stewart theory. In both references it is studied the thermodynamic consistency of the phantom solution through the first and second derivative of the entropy, resulting that in the non-linear case this consistency can be guaranteed only in a small region of the space of parameters ($\omega, \xi_{0}$).\\ 

After performing a direct integration of the solution (\ref{eq:phantom}) we can obtain
\begin{equation}
a(t) = a(t_{0})\left(\frac{t_{s}-t}{t_{s}-t_{0}} \right)^{-A},
\end{equation}
and also we can have an expression for the time at which the big rip will occur in terms of the solution $A$, $t_{s} = t_{0} + A/H(t_{0})$. For practical purposes the constant $t_{0}$ will represent some initial time. For the case when $t = t_{s}$, the scale factor diverges. Using the expression (\ref{eq:phantom}) in the equation (\ref{eq:pressure}) one gets
\begin{equation}
\Pi(t) = -\left[2A+3(1+\omega)A^{2} \right](t_{s}-t)^{-2},
\label{eq:dissipation}
\end{equation}
which exhibits a divergent behavior as $t \rightarrow t_{s}$, therefore the effective pressure diverges. By means of Eq. (\ref{eq:fried1}) we can write the density of the dissipative fluid as $\rho(t) = 3A^{2}(t_{s}-t) ^{-2}$, that also is divergent when $t=t_{s}$, according to these characteristics the solution (\ref{eq:phantom}) represents a Type I singularity, i.e., a big rip \cite{odintsov}.

\subsection{Little Rip Cosmology}
As pointed out in Ref. \cite{nojiri} it is possible that phantom regime is only a transient stage of the universe and there may be various mechanisms that could prevent a future singularity. There exists an interesting scenario where the scale factor and the energy density do not diverge at a finite future time, however, such model can drive to structure disintegration, this is the little rip cosmology \cite{little0}. This scheme could be a viable alternative to the $\Lambda$CDM model. In Ref. \cite{little} it was found that the inclusion of viscous effects in a standard fluid in the following form    
\begin{equation}
p_{eff} = p + \Pi = p - 3\xi H,
\end{equation} 
can lead to a little rip evolution. The previous expression for the bulk viscous pressure can be obtained from Eq. (\ref{eq:transport}) when $\tau \rightarrow 0$. In general grounds, an exploration of this scenario could be relevant if it allows us to establish bounded quantities at thermodynamic level.

\subsection{Cosmological perturbations}
Different dark energy models having a phantom behavior can show a distinct future evolution, i.e., big rip singularity, little rip cosmology, pseudo-rip universe, among other final states. Each of these final states are characterized by the critical behavior of a specific cosmological quantity at the moment of the singularity (scale factor, energy density of the fluid, derivatives of the scale factor, etc.), however in the literature can be found that several dark energy models can mimic the $\Lambda$CDM model, i.e., the parameter state of the models are close to the value $-1$ but below this. In Ref. \cite{pert1} can be found that for several dark energy models with a future singularity, the growth of the density and matter perturbations is similar as the one obtained for the $\Lambda$CDM model, but with the difference that an instant before the singularity such perturbations become large and this occurs before possible disintegration of bound structures. The final state of the universe results to be chaotic. The complete set of equations for cosmological perturbations for these dark energy models is given by
\begin{align}
& \delta'_{DE} + 3(1-\omega_{DE})aH\delta_{DE}-(1+\omega_{DE})\delta_{m} \nonumber \\ 
& + (1+\omega_{DE})\left(k+9a^{2}H^{2}\frac{1-c^{2}_{a}}{k}\right)V_{DE}=0,\\
& V'_{DE}-2aHV_{DE}-\frac{k}{1+\omega_{DE}}\delta_{DE} = 0,\\
& \delta''_{m}+\frac{a'}{a}\delta'_{m}-\frac{1}{2}(\rho_{m}\delta_{m}+(1+3\omega_{DE})\rho_{DE}\delta_{DE}) = 0,
\end{align}
where the prime denotes a derivative with respect to a conformal time, $\delta_{DE} = \delta \rho_{DE}/\rho_{DE}$, $V_{DE}$ is the velocity perturbation of dark energy, $c^{2}_{a}$ is the adiabatic speed of sound and $\delta_{m}$ are the matter perturbations. By $k$ the wavenumber of the corresponding mode is denoted. See also the reference \cite{pert2}, where the future evolution of dark energy density-matter perturbations were described. The perturbations for this model are also solved by the previous system of equations. In this work it results that in a scalar field model there exists a slow evolution from $\Lambda$CDM to a big rip final state or in other words, the growth of the perturbations becomes relevant only at the singularity. In Ref. \cite{maartens} it can be found that the Israel-Stewart model is stable under linear perturbations and besides in Ref. \cite{ccl} it was shown also in the framework of causal Israel-Stewart model that a solution-type as the one given in (\ref{eq:phantom}) presents stability along the cosmic evolution.           

\section{Irreversible thermodynamics}
\label{sec:thermo}  
\begin{figure}[htbp!]
\centering
\includegraphics[scale=0.65]{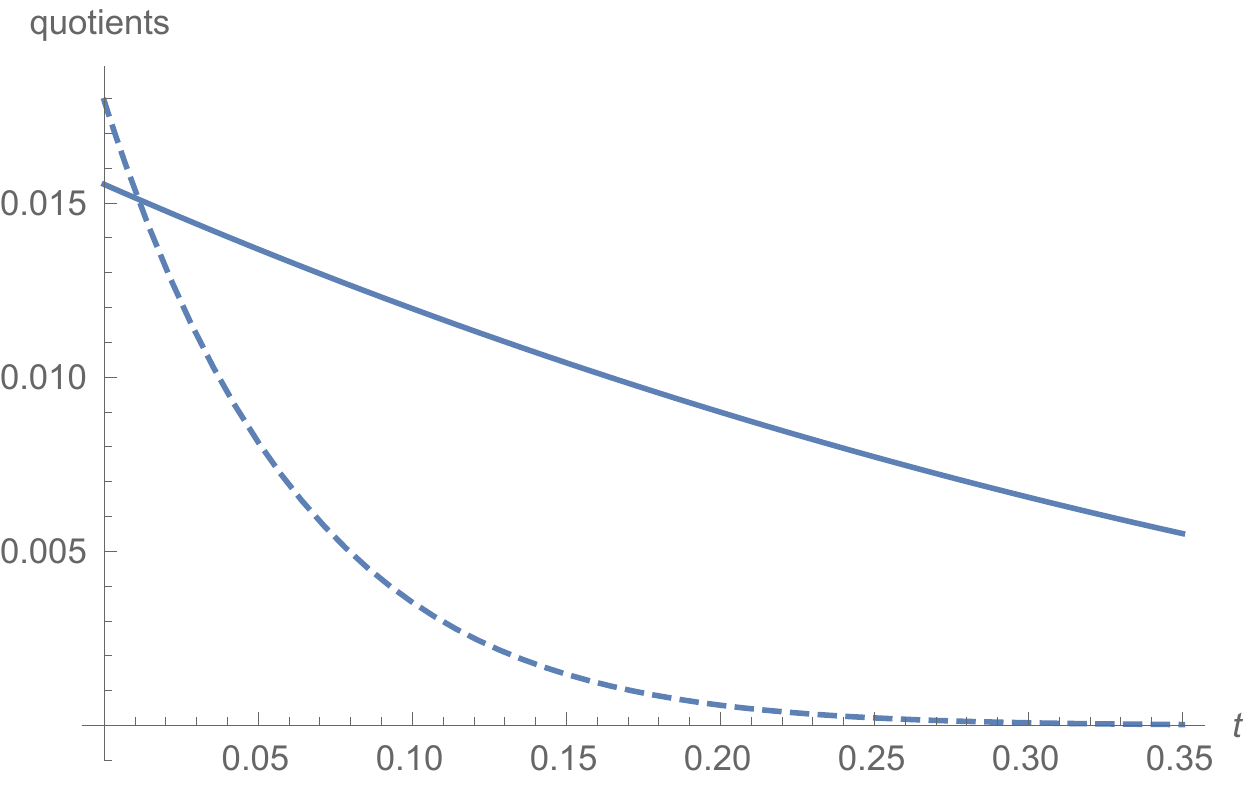}
\caption{The solid line represents the quotient $n(t)/\left| \Pi(t)\right|$ and the dashed line $n(t)/\rho(t)$.} 
\label{fig:quot}
\end{figure}

Perfect fluids in equilibrium do not generate entropy and dissipation processes are not included in their description, but as far as we know real fluids behave irreversibly, therefore we must consider a relativistic theory of dissipative fluids. For a homogeneous and isotropic universe, the evolution equation for the entropy will be given by \cite{maartens2}
\begin{equation}
nT\frac{dS}{dt} = -3H\Pi,
\label{eq:entropy}
\end{equation}
where $n$ is the number density of particles. As can be seen, for an expanding universe the entropy production will be positive if $\Pi < 0$. The inclusion of dissipative processes leads to a non-conservation of the particle number, then
\begin{equation}
\frac{\dot{n}}{n} + 3H = \frac{\dot{N}}{N},
\label{eq:production}
\end{equation}\\
where the quantity $\dot{N}/N := \nu$ is the production rate of particles and $N := nV$, being $V$ the volume containing the $N$ particles. For instance, if we consider a constant production rate, $\nu$, one gets
\begin{equation}
N(t) = N(t_{0})e^{\nu (t-t_{0})},
\label{eq:number}
\end{equation}
then, inserting this result in Eq. (\ref{eq:production}) we obtain
\begin{equation}
n(t) = n(t_{0})\left(\frac{a(t_{0})}{a(t)} \right)^{3}e^{\nu \left(t-t_{0}\right)}.
\label{eq:particles}
\end{equation} 
On the other hand, if we consider the Eqs. (\ref{eq:phantom}) and (\ref{eq:number}) in the expression (\ref{eq:production}) we obtain
\begin{equation}
n(t) = n(t_{0})\left(\frac{t_{s}-t}{t_{s}-t_{0}} \right)^{3A}e^{\nu \left(t-t_{0}\right)},
\label{eq:producidas}
\end{equation}
implying that the number density of particles tends to zero as $t \rightarrow t_{s}$. This result is consistent since as the universe approaches the singularity the number of particles reaches a fixed value given by $N(t_{s}) = N(t_{0})e^{\nu (t_{s}-t_{0})}$, but the volume increases indefinitely. It is worthy to mention that this behavior is independent of the production or decay rate of particles. If we consider the Eqs. (\ref{eq:dissipation}) and (\ref{eq:producidas}) we can construct the quotient $n(t)/\left| \Pi(t)\right| \propto (t_{s}-t)^{3A + 2}$, also we can compute the quotient $n(t)/\rho(t)$, which has the same behavior as the previous quotient as a function of time. In FIG. \ref{fig:quot} we can observe the behavior of both quotients. The value of $A$ corresponds to the solution found in \cite{lepe1} with $\omega = 0.0001$, for the plot we considered $\nu = -1/2$, $t_{0} = 0$, $t_{s} = 1$ and $n(t_{0}) = 1$. From this quotients we can infer that the dissipative effects annihilate the particles of the phantom fluid but at the same time act like energy density catalyst. The quotient $\rho(t)/\left| \Pi(t)\right| = \mbox{constant}$.  
\\

According to the second law of thermodynamics, we can write the following expression \cite{callen}
\begin{equation}
TdS = d(\rho V) + pdV - \mu dN,
\label{eq:euler}
\end{equation}
where $\rho V$ is the internal energy and $\mu$ is the chemical potential, we will also consider $V$ as the Hubble volume given as $V(a)= V(a_{0})(a/a_{0})^{3}$ in this form $dV/V = 3(a/a_{0})^{-1}d(a/a_{0}) = 3Hdt$, which can be written as $\dot{V}/V = 3H$. If we take the cosmic time derivative of equation (\ref{eq:euler}) and deal with a barotropic equation of state for the density and pressure of the fluid, we obtain
\begin{equation}
\dot{\rho} + \rho(1+\omega)\frac{\dot{V}}{V} = \frac{1}{V}\left(\mu \dot{N}+T\frac{dS}{dt} \right).
\end{equation}
Using the relation $\dot{V}/V = 3H$ and the Eqs. (\ref{eq:continuity}), (\ref{eq:entropy}) in the preceding expression, one gets
\begin{equation}
\mu n \frac{\dot{N}}{N} = \mu n \nu = -3H\Pi \left(1-\frac{1}{N}\right),
\end{equation} 
where we used $N = nV$ and the definition of the production rate of particles, therefore
\begin{equation}
\mu = -\frac{3H}{n \nu}\left(\frac{N-1}{N}\right)\Pi.
\label{eq:newchemical}
\end{equation}
As can be seen in the previous equation, unlike what it is considered in the reference \cite{lima1}, where the negative chemical potential must be introduced by hand, we can see that the framework of irreversible thermodynamics allows us to have a negative chemical potential. Finally, if we insert the phantom solution (\ref{eq:phantom}) together with the Eqs. (\ref{eq:dissipation}) and (\ref{eq:producidas}) in the expression for the chemical potential given in (\ref{eq:newchemical}), results
\begin{eqnarray}
\mu(t) &=& \frac{3A^{2}}{n(t_{0})\nu}\frac{(t_{s} - t_{0})^{3A}}{e^{\nu(t-t_{0})}}\left(\frac{N-1}{N}\right)\times \nonumber \\
& \times & \left[2+3(1+\omega)A\right](t_{s}-t)^{-3(1+A)}.
\label{eq:chemical}
\end{eqnarray}
If we consider a negative production rate of particles, $\nu < 0$, the chemical potential (\ref{eq:chemical}) will be negative. According to the value taken by the parameter $\nu$, we will have production or annihilation of particles, then, this is an agreement with the thermodynamics point of view, the chemical potential acts like a {\it generalized force} for matter flow between two interacting systems \cite{callen}. The particle production (or decay) process in the universe has been widely studied and it has to be approached from a quantum perspective \cite{birrel}. Besides, in Ref. \cite{zeldovich} it was pointed out that non-conservation of the particle production number in cosmology can be due to the presence of {\it viscosity} in the vacuum. Also, in Ref. \cite{triginer} it was shown that a effective viscous pressure approach is compatible with the kinetic theory of a Maxwell-Boltzmann gas with non-conserved particle number. For the limit case, $t \rightarrow t_{s}$, the chemical potential given in Eq. (\ref{eq:chemical}) exhibits a divergent behavior.

\section{Temperature and Entropy}
\label{sec:temp}
In order to determine the effect of the chemical potential on the thermodynamics of the model, we must consider the Euler relation, which can be obtained from the expression (\ref{eq:euler})
\begin{equation}
TS = (1+\omega)\rho V - \mu N,
\end{equation}
and we have considered a barotropic equation of state. Under the description of the standard thermodynamics for a homogeneous fluid, the temperature is always positive definite, therefore $TS > 0$ implying that $S > 0$ \cite{lima1}. Note that the aforementioned condition for $TS$ could not be guaranteed in the phantom regime ($\omega < -1$). See for instance, if $\omega < -1$, we have
\begin{equation}
TS = -\left(\left|1+\omega \right| \rho V + \mu N \right) < 0, 
\end{equation}   
then $T < 0$ and $S > 0$ or vice versa, however, either of the two cases involves problems at thermodynamic level. Based on the results of irreversible thermodynamics for a dissipative fluid, we have the following equation for the temperature evolution \cite{maartens2}
\begin{equation}
\frac{\dot{T}}{T} = \frac{\omega \dot{\rho}+3H\Pi}{(1+\omega)\rho}\left[1+\frac{3H\Pi}{nT} \right],
\end{equation}
which can be rewritten as follows
\begin{align}
& \frac{dT}{dt} = 3H\left\lbrace \omega + \frac{(1-\omega)}{(1+\omega)}\left[(1+\omega)+\frac{2\dot{H}}{3H^{2}}\right]\right\rbrace \times \nonumber \\
& \times \left\lbrace \frac{9H^{3}}{n}\left[(1+\omega)+\frac{2\dot{H}}{3H^{2}} \right]-T \right\rbrace,
\label{eq:temperature}
\end{align}
where we have used the Eqs. (\ref{eq:fried1}), (\ref{eq:continuity}) and (\ref{eq:pressure}). If we consider the phantom solution (\ref{eq:phantom}) in the above equation, we have
\begin{align}
& dT = \mathbf{C}_{1}\left(\mathbf{C}_{2}\frac{(t_{s}-t)^{-3}}{n(t)}-T \right)\frac{dt}{(t_{s}-t)}, \nonumber\\ 
& = \mathbf{C}_{1}\left(T - \frac{\mathbf{C}_{2}(t_{s}-t)^{-3(A+1)}}{n(t_{0})(t_{s}-t_{0})^{-3A}}e^{-\nu(t-t_{0})}\right) d\ln (t_{s}-t). 
\label{eq:dertemp}
\end{align}
In the last expression we have considered the expression (\ref{eq:producidas}) for $n(t)$ and for simplicity in the notation we have defined the following two positive constants
\begin{eqnarray*}
\mathbf{C}_{1} &=& 3A\left\lbrace \omega + \frac{(1-\omega)}{(1+\omega)}\left[(1+\omega)+\frac{2}{3A}\right]\right\rbrace, \\
\mathbf{C}_{2} &=& 9A^{3}\left[(1+\omega)+\frac{2}{3A}\right].
\end{eqnarray*}
After a straightforward calculation, the expression (\ref{eq:dertemp}) can be simplified to the form
\begin{equation}
\frac{dT}{d\ln (t_{s}-t)} = \mathbf{C}_{1}T - \mathbf{C}_{1}\mathbf{C}_{3}(t_{s}-t)^{-3(A+1)}e^{\nu(t_{s}-t)},
\label{eq:tempbuena}
\end{equation}
where $C_{3}$ is a positive constant given as follows
\begin{equation*}
\mathbf{C}_{3} = \frac{\mathbf{C}_{2}e^{-\nu(t_{s}-t_{0})}}{n(t_{0})(t_{s}-t_{0})^{-3A}}.
\end{equation*}
Finally, by considering the change of variable $x := t_{s}-t$, the equation for the temperature evolution (\ref{eq:tempbuena}) can be expressed as the following differential equation
\begin{equation}
\frac{dT}{dx}-\frac{\mathbf{C}_{1}}{x}T + \mathbf{C}_{1}\mathbf{C}_{3}x^{-3(A+1)-1}e^{\nu x} = 0,
\end{equation}
which has analytical solution given as
\begin{widetext}
\begin{equation}
T(t) = \mathcal{C}(t_{s}-t)^{\mathbf{C}_{1}} + \mathbf{C}_{1}\mathbf{C}_{3}(\nu \left[-(t_{s}-t)\right])^{\mathbf{C}_{1}+3(A+1)}(t_{s}-t)^{-3(A+1)}\Gamma(-[\mathbf{C}_{1}+3(A+1)], -\nu (t_{s}-t)),
\label{eq:temptiempo}
\end{equation}
\end{widetext}
where $\mathcal{C}$ is an integration constant and $\Gamma(a,x)$ is the incomplete gamma function. If we consider $T(t=t_{0})$, the quotient $T(t)/T(t_{0})$ takes the form
\begin{widetext}
\begin{equation}
\frac{T(t)}{T(t_{0})} =\frac{\mathcal{C}\left(1-\frac{t}{t_{s}}\right)^{\mathbf{C}_{1}} + \mathbf{C}_{1}\mathbf{C}_{3}(\nu \left[-\left(1-\frac{t}{t_{s}}\right)\right])^{\mathbf{C}_{1}+3(A+1)}\left(1-\frac{t}{t_{s}}\right)^{-3(A+1)}\Gamma \left(-[\mathbf{C}_{1}+3(A+1)], -\nu t_{s}\left(1-\frac{t}{t_{s}}\right)\right)}{\mathcal{C}\left(1-\frac{t_{0}}{t_{s}}\right)^{\mathbf{C}_{1}} + \mathbf{C}_{1}\mathbf{C}_{3}(\nu \left[-\left(1-\frac{t_{0}}{t_{s}}\right)\right])^{\mathbf{C}_{1}+3(A+1)}\left(1-\frac{t_{0}}{t_{s}}\right)^{-3(A+1)}\Gamma \left(-[\mathbf{C}_{1}+3(A+1)], -\nu t_{s}\left(1-\frac{t_{0}}{t_{s}}\right)\right)},
\label{eq:temptiempo2}
\end{equation}
\end{widetext}
where we can identify an explicit form for $T(t_{0})$.\\ 

\begin{figure}[htbp!]
\centering
\includegraphics[scale=0.65]{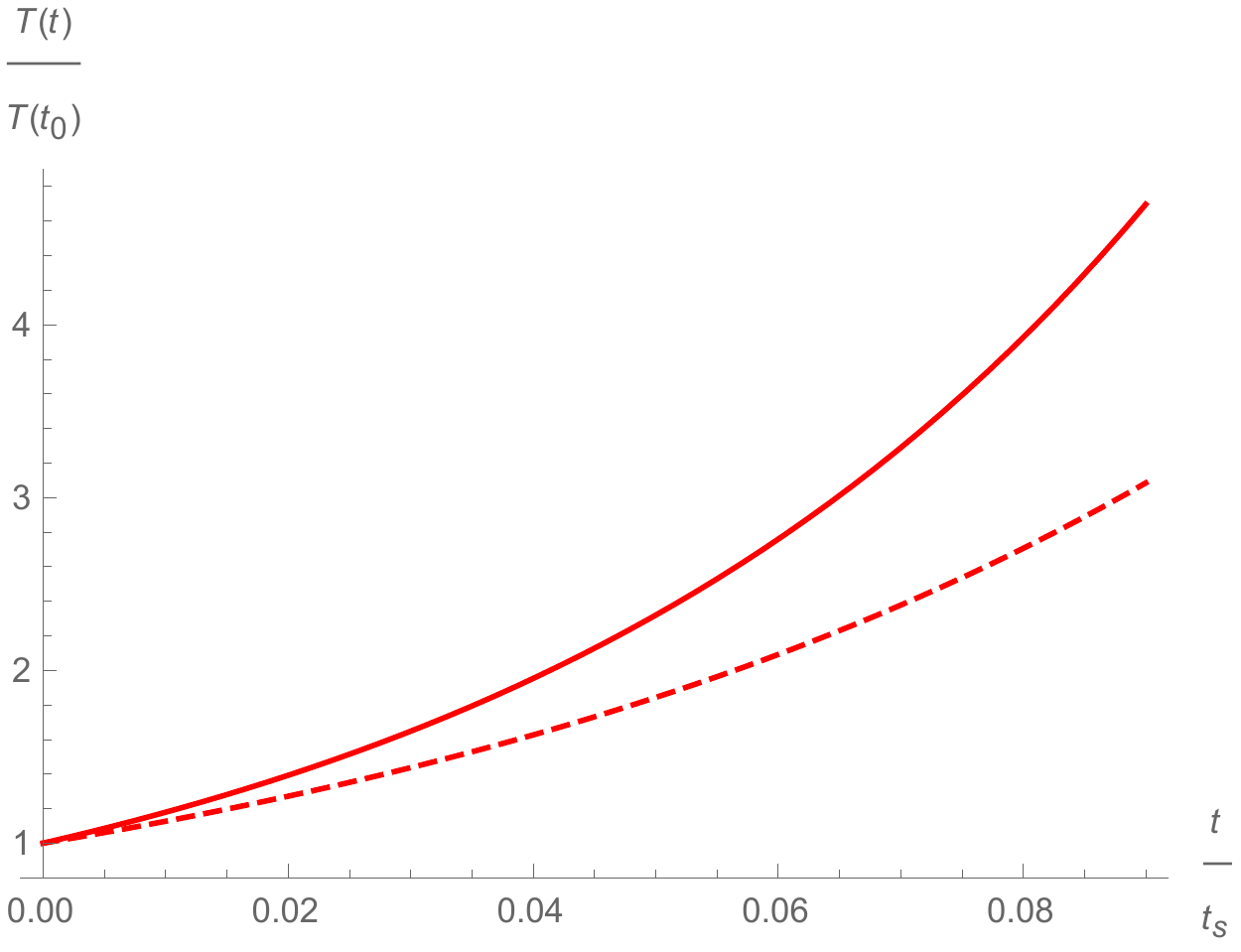}
\caption{Behavior of the temperature as a function of time. The solid line was obtained with $\xi_{0} = 2$ and the dashed line with $\xi_{0} = 3$.} 
\label{fig:temperature}
\end{figure}

In FIG. \ref{fig:temperature} we can observe the behavior of the quotient given in (\ref{eq:temptiempo2}) as a function of time. For this plot we have considered the positive solution for $A$ obtained in Ref. \cite{lepe1} with $\epsilon = 1$ and $\omega = 0.0001$, for other values of $\epsilon$ we obtain a similar behavior as the one shown in the FIG. \ref{fig:temperature}. The parameter $\epsilon$ is rectricted to the interval $(0,1]$ for causality reasons. For simplicity we also consider $t_{0} = 0$ and $\nu = -1/2$. For initial number density of particles we considered, $n(t_{0}) = 1$. On the other hand, as $t/t_{s} \rightarrow 1$ the value of the temperature diverges, but keeps positive along the evolution. As can be observed in the plot, as the value of the parameter $\xi_{0}$ increases, the growth of the temperature is slower. The values considered for the constant $\xi_{0}$ were chosen in the interval for which $A > 0$ since we are interested in the description of an expanding universe \cite{lepe1}. A similar behavior as shown in FIG. \ref{fig:temperature} is obtained for other small values of the parameter $\omega$.
\\

\begin{figure}[htbp!]
\centering
\includegraphics[scale=0.65]{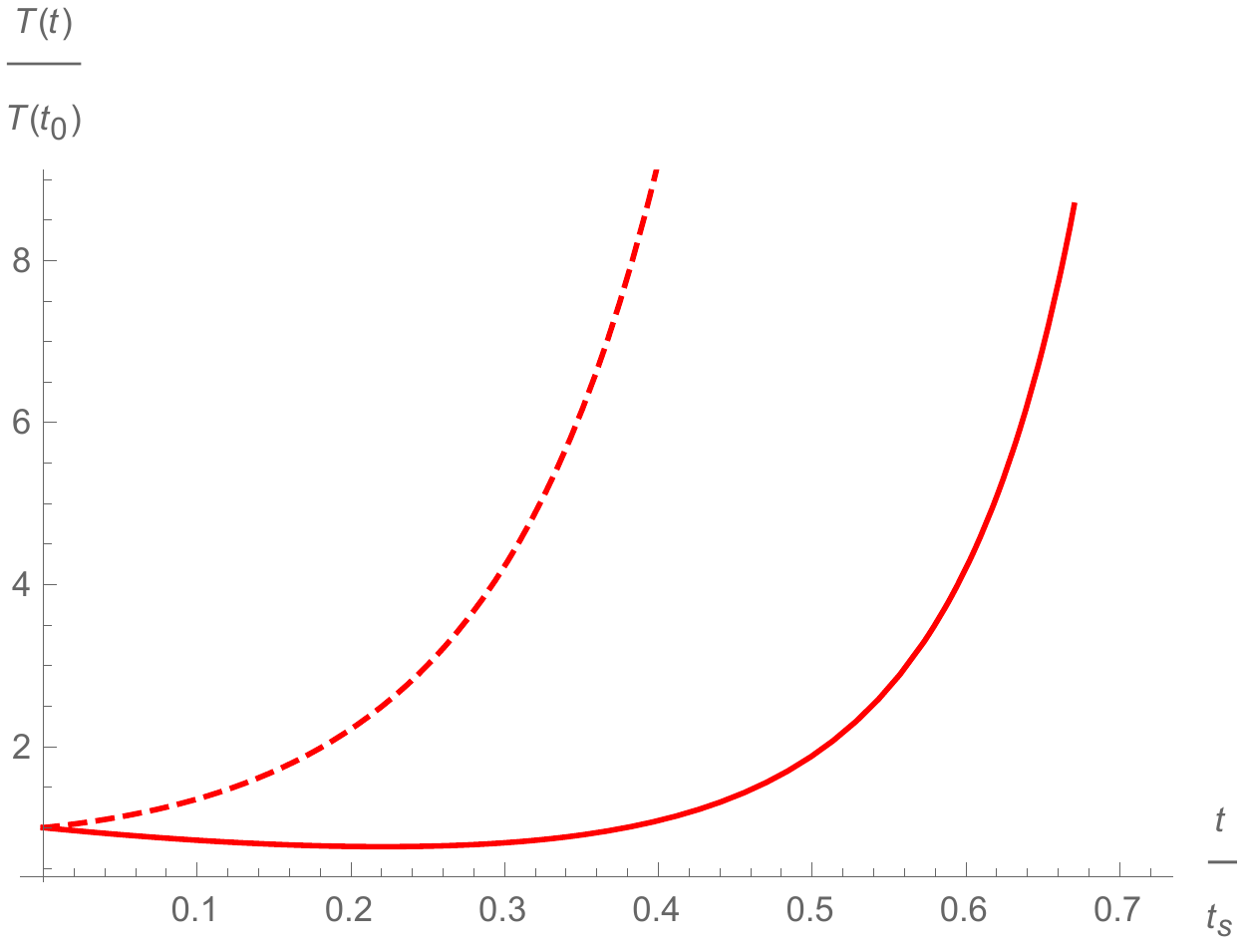}
\caption{Behavior of the temperature as a function of time in the non-linear extension of the Israel-Stewart model. The solid line was obtained with $\xi_{0} = 0.18$ and the dashed line with $\xi_{0} = 0.7$.} 
\label{fig:temperature1}
\end{figure}

In FIG. \ref{fig:temperature1}, we depict the growth of the temperature by considering the corresponding values of the solution $A$ found in the non-linear extension of the Israel-Stewart model \cite{lepe2}. As in the previous case we have $\omega = 0.0001$, $t_{0} = 0$, $\nu = -1/2$ and for the initial number density of particles we considered, $n(t_{0}) = 1$. It is important to point out that in this case the growth of the temperature is slower than in the Israel-Stewart model and contrary to the Israel-Stewart framework, as the value of the parameter $\xi_{0}$ increases, the growth of the temperature is faster.   
\\

The phantom behavior for the Israel-Stewart model and also in a non-linear extension of this is guaranteed always that the positivity of the parameter $\xi_{0}$ is fulfilled, but, in Ref. \cite{negativebulk} it can be found that the consideration of negative values for $\xi_{0}$ can lead to a parameter state below -1 and positive temperature, however the viscosity contributes with attractive gravity, i.e., the cosmic expansion is not allowed in this description and the model presents other fundamental problems at thermodynamics level.\\
 
It is worthy to mention that the positivity of the temperature depends on the value taken by the integration constant, $\mathcal{C}$, if we evaluate the expression (\ref{eq:temptiempo2}) at $t/t_{s} \ll 1$ and consider that the value of the quotient must be positive, we obtain the following condition for $\mathcal{C}$
\begin{equation}
\mathcal{C} > -(-1)^{\mathbf{C}_{1}+3(A+1)}\mathbf{C}_{1}\mathbf{C}_{3}\nu^{\mathbf{C}_{1}+3(A+1)}\Gamma(-[\mathbf{C}_{1}+3(A+1)]),
\label{eq:conTemp}
\end{equation}
where $\Gamma(a)$ is simply the gamma function. As can be seen from the above equation, the integration constant $\mathcal{C}$ is dependent of the production rate of particles, $\nu$ (which is considered as constant), and of the parameters $(\omega, \xi_{0})$. For a negative production rate of particles the condition given in (\ref{eq:conTemp}) for $\mathcal{C}$ can be simplified to 
\begin{equation}
\mathcal{C} > - \mathbf{C}_{1}\mathbf{C}_{3}\nu^{\mathbf{C}_{1}+3(A+1)}\Gamma(-[\mathbf{C}_{1}+3(A+1)]).
\label{eq:conTemp1}
\end{equation}
Besides, the resulting value for $\mathbf{C}_{1}+3(A+1)$ must not be an integer number in order to avoid the poles of the gamma function. Finally, in both FIGS. \ref{fig:temperature}-\ref{fig:temperature1} we can observe that $[T(t/t_{s} \ll 1)/T(t_{0})] \neq 0$ for all the considered cases, note that in this limit the solution (\ref{eq:phantom}) is given by a constant value $H(t/t_{s} \ll 1) \rightarrow A/t_{s}$, i.e., for a de Sitter type evolution the temperature takes a fixed value.\\

To end this section, using all the previous results, the entropy resulting from the equation (\ref{eq:entropy}) is positive definite
\begin{equation}
\frac{dS}{dt} = \frac{3H}{T}\frac{\left| \Pi\right|}{n} \propto \frac{1}{T}(t_{s}-t)^{-3(1+A)},
\end{equation}
on the other hand, by taking the time derivative of the previous expression one gets
\begin{equation}
\frac{d^{2}S}{dt^{2}} \propto -\left[3(1+A)\frac{1}{T}(t_{s}-t)^{-1}+\frac{1}{T^{2}}\frac{dT}{dt}\right](t_{s}-t)^{-3(1+A)},
\end{equation}
therefore the consistency thermodynamics conditions for the entropy: $dS/dt > 0$ and $d^{2}S/dt^{2} < 0$ are satisfied for this model in the phantom regime.
\section{Final remarks} 
\label{sec:elfinal}
The present work was devoted to study the definiteness of the phantom regime with a solution given in the following form for the Hubble parameter, $H(t) = A(t_{s}-t)^{-1}$, from a thermodynamics point of view in an isotropic and homogeneous cosmology and when dissipative effects are taken into account. This was done by considering a causal description for the dissipative effects given by the full Israel-Stewart transport equation, a non-zero chemical potential, $\mu$, the thermodynamics of irreversible processes and non-conserved particle number in the cosmological fluid. We have found that with these considerations we can construct a chemical potential that results to be negative.\\
It is important to point out the if we consider annihilation of particles ($\nu < 0$) in the fluid once the phantom solution is inserted in the expression for the chemical potential, this will be always negative and can be interpreted as a {\it generalized force} for matter flow, since the production or decay rate of particles is equivalent to a bulk viscous pressure in the cosmological fluid.\\

By considering the evolution equation for the temperature of a dissipative fluid, we can see that the resulting temperature is positive throughout the cosmic evolution, with an initial value different from zero. This initial value can be modulated by the integration constant $\mathcal{C}$ and we could establish a minimum value for this constant in order to maintain a positive temperature. This is in agreement with the standard thermodynamics of an homogeneous fluid. It is worthy to mention that since we are considering very small values for the $\omega$ parameter, it means that the matter present in the cosmic fluid is dark matter type with dissipative effects. As is well known, the so-called WIMPs are candidates for dark matter and their detection until now remains as a challenge. A proposal for the possible detection of these particles is given by the construction of detectors that operate at very low temperatures (below $1$ K), however there has been no success in this area, it is expected that the sensitivity of these detectors could be improved in future experiments \cite{detection}, note that dissipative effects, if they exist, could contribute significantly in the detection of these particles through their temperature.\\
 
Finally, using the result obtained for the temperature and the evolution equation for the entropy, we can see that $S > 0$ for an expanding universe. Therefore, the positivity for both quantities, entropy and temperature, can be guaranteed at same time. Hence, at first time it is proposed the resolution of negative entropy or negative temperature problem in accelerating phantom universe.
\\

In summary, the inclusion of dissipative effects described by the thermodynamics of irreversible processes in a cosmological fluid, allows us to have access to the phantom zone without entering into contradictions with the definition of temperature and entropy given by standard thermodynamics.

\section*{Acknowledgments}
M.C. work is supported by S.N.I. (CONACyT-M\'exico) and PRODEP (UV-PTC-851). S.D.O. work is supported by MINECO (Spain), FIS2016-76363-P, by project 2017 SGR247 (AGAUR, Catalonia) and by Russian Min. of Education and Science, project No. 3.1386.2017.

\end{document}